\documentclass[preprint2]{aastex}

\def \h2         {\hbox{H$_2$}}

\def\approxlt{\lower.2em\hbox{$\buildrel < \over \sim$}}
\def\approxgt{\lower.2em\hbox{$\buildrel > \over \sim$}}

\begin{document}

\title {Interferometric Follow-Up of WISE Hyper-Luminous Hot, Dust-Obscured Galaxies}

\author {Jingwen Wu\altaffilmark{1},  
R. Shane Bussmann\altaffilmark{2,3},
Chao-Wei Tsai\altaffilmark{4,5}, 
Andreea Petric\altaffilmark{6},
Andrew Blain\altaffilmark{7},
Peter R. M. Eisenhardt\altaffilmark{4},
Carrie R. Bridge\altaffilmark{8},
Dominic J. Benford\altaffilmark{9},
Daniel Stern\altaffilmark{4},
Roberto J. Assef\altaffilmark{10}, 
Christopher R. Gelino\altaffilmark{11},
Leonidas Moustakas\altaffilmark{4},
Edward L. Wright\altaffilmark{1}
}

\altaffiltext{1}{Department of Physics and Astronomy, University of California, Los Angeles, CA 90095, USA; jingwen@astro.ucla.edu}
\altaffiltext{2}{Harvard-Smithsonian Center for Astrophysics, 60 Garden St. MS78, Cambridge, MA, 02138, USA}
\altaffiltext{3}{Department of Astronomy, Space Sciences Building, Cornell University, Ithaca, NY 14853, USA}
\altaffiltext{4}{Jet Propulsion Laboratory, California Institute of Technology, 4800 Oak Grove Dr., Pasadena, CA 91109, USA}
\altaffiltext{5}{NASA Postdoctoral program Fellow}
\altaffiltext{6}{Institute for Astronomy, 2680 Woodlawn Drive, Honolulu, HI 96822-1839, USA}
\altaffiltext{7}{Department of Physics and Astronomy, University of Leicester,  Leicester, LE1 7RH, UK}
\altaffiltext{8}{Division of Physics, Math and Astronomy, California Institute of Technology, Pasadena, CA 91125, USA}
\altaffiltext{9}{NASA Goddard Space Flight Center, Greenbelt, MD 20771, USA}
\altaffiltext{10}{N\'ucleo de Astronom\'ia de la Facultad de Ingenier\'ia, Universidad Diego Portales, Av. Ej\'ercito Libertador 441, Santiago, Chile.}
\altaffiltext{11}{Infrared Processing and Analysis Center, California Institute of Technology, Pasadena, CA 91125, USA}

\begin{abstract}

WISE has discovered an extraordinary population of hyper-luminous
dusty galaxies which are faint in the
two bluer passbands ($3.4\, \mu$m and $4.6\, \mu$m) but are
bright in the two redder passbands of WISE ($12\, \mu$m and $22\,
\mu$m).  We report on initial follow-up observations of three of these hot,
dust-obscured galaxies, or Hot DOGs, using the CARMA and SMA
interferometer arrays at submm/mm wavelengths.  We report continuum
detections at $\sim$ 1.3 mm of two sources (WISE J014946.17+235014.5 and
WISE J223810.20+265319.7, hereafter W0149+2350 and W2238+2653, respectively), and upper limits to CO line emission at 3 mm in the observed frame for two sources (W0149+2350 and WISE J181417.29+341224.8, hereafter W1814+3412). The 1.3 mm continuum images have a
resolution of $1-2$\arcsec\ and are consistent with single point sources. We estimate the masses of cold dust are 2.0$\times 10^{8} M_{\odot}$ for W0149+2350 and  3.9$\times 10^{8} M_{\odot}$ for W2238+2653, comparable to cold dust masses of luminous quasars. We obtain 2$\sigma$ upper limits to the molecular gas masses traced by CO, which are  3.3$\times 10^{10} M_{\odot}$ and  2.3$\times 10^{10} M_{\odot}$ for W0149+2350 and W1814+3412, respectively.
We also present high-resolution, near-IR imaging with
WFC3 on the {\it Hubble Space Telescope} for W0149+2653 and with
NIRC2 on Keck for W2238+2653. The near-IR images show morphological
structure dominated by a single, centrally condensed source with
effective radius less than 4 kpc. No signs of gravitational lensing are
evident.

\end{abstract}

---galaxies: individual (WISE J014946.17+235014.5, WISE J181417.29+341224.9,  WISE J223810.20+265319.7)

\section{Introduction} \label{intro}

Interferometric observations at submm/mm bands play a unique role in exploring the nature of dusty galaxies, particularly at $z \sim2-3$, corresponding to the peak epoch of cosmic star formation and quasar activity. Lower-J CO transitions (e.g., CO 3-2 and CO 4-3), redshifted to $\sim3$ mm at $z \sim2-3$, act as tracers of the overall molecular gas reservoir, while the continuum emission near observed-frame 1-3 mm provides reliable  constraints on cold dust masses. High-resolution CO and continuum observations at 1-3 mm provide crucial information to study the gas content, morphology, and dynamics of dusty galaxies. 

Great progress has been made in the last decade in the study of cool gas and dust in distant galaxies (e.g., Solomon \& Vanden Bout et al. 2005, Carilli \& Walter 2013). Close to 200 galaxies at $z >$1 have been observed for molecular gas, most with interferometric telescopes, and many with longer wavelength (e.g., $\sim$ 1 mm) continuum detected (see Carilli \& Walter 2013, and references therein).  There seem to be two parallel sequences of star forming galaxies based on their cool interstellar medium (ISM)  properties (e.g., Dav{\'e} et al. 2010, Engel et al. 2010, Carilli \& Walter 2013): ``main sequence" galaxies, which follow the normal star formation rate $-$ stellar mass correlation as in the Milky Way, including normal spiral galaxies in the local universe, color-selected star-forming galaxies (CSGs, e.g., Carilli \& Walter 2013) at $z$=1.5 to 3, and Lyman Break Galaxies (LBGs) at $z >$3; and the higher-efficiency off-main sequence starburst galaxies (hereafter  ``bursting galaxies"), including local starburst galaxies, submillimeter galaxies (SMGs -- Blain et al. 2002, Chapman et al. 2005), {\it Spitzer} 24 $\mu$m-selected dust-obscured galaxies (DOGs --Dey at al. 2008), optically selected quasars (QSOs), and radio galaxies (RGs). According to some popular theoretical models of galaxy evolution through mergers (e.g., Hopkins et al. 2008), galaxies outside of the main sequence may correspond to interesting unusually high-luminosity episodes in galaxy evolution. SMGs likely represent an early phase in galaxy-galaxy merging when gas is funneled into the center of galaxies, triggering starbursts, which overwhelm the bolometric luminosity (e.g., Men{\'e}ndez-Delmestre et al. 2009). Then a central supermassive black hole (SMBH) is fueled, changing the major power source from the starburst to the AGN as the galaxy goes through the DOG phase (e.g., Dey et al. 2008, Narayanan et a;. 2010). Finally, AGN feedback quenches further star formation and evacuates surrounding dust, leaving an optically bright quasar (e.g., Hopkins et al. 2008). 

Recently, a new population of $z \sim$ 2-3 galaxies was discovered (Eisenhardt et al. 2012, Wu et al. 2012, Bridge et al. 2013) using NASA's Wide-field Infrared Survey Explorer (WISE) mission (Wright et al. 2010).
By selecting targets that are prominent in the WISE 12 $\mu$m (W3) or 22 $\mu$m (W4) bands, and faint or undetected in the 3.4 $\mu$m (W1) and 4.6 $\mu$m (W2) bands, Eisenhardt et al. (2012) identified an all-sky sample of about 1000 ``W1W2-dropout'' objects. Spectroscopic follow-up observations reveal that most of these targets are at redshift 1 to 4, with a peak at $z\sim 2-3$ similar to what is observed for SMGs and DOGs. Many W1W2-dropout spectra show obscured AGN features (Wu et al. 2012, Eisenhardt et al. in prep). Submillimeter follow-up observations using SHARC2 and Bolocam at the Caltech Submillimeter Observatory (CSO) (Wu et al. 2012), SCUBA2 at the James Clerk Maxwell Telescope (JCMT) (Jones et al. 2014), and PACS and SPIRE on {\it Herschel} (Bridge et al. 2013, Tsai et al. in prep) reveal that they are hyper-luminous: most have luminosities well over 10$^{13}$\,$L_\odot$, 
and some exceed 10$^{14}$\,$L_\odot$, comparable to the most luminous quasars known (Tsai et al. in prep., Assef et al. in prep.). However, their SEDs are distinct from other known populations, with an unusually high mid-IR to submm luminosity ratio, suggesting their luminosities are dominated by emission from hot dust. These galaxies also satisfy the selection criteria for DOGs, which are optically faint, and 24\,$\mu$m bright, corresponding to $F_{24}/F_{R} > 1000$ (Dey et al. 2008). However, compared to DOGs identified in the 9 deg$^{2}$ Bo$\ddot{\rm o}$tes field (e.g., Dey et al. 2008, Melbourne et al. 2012), the WISE-selected sources are at least 2 times hotter, 10 times more luminous, and 10,000 times rarer. Therefore, Wu et al. (2012) refer to these galaxies as Hot Dust-Obscured Galaxies or ``Hot DOGs."

Hot DOGs are a rare population with extremely high luminosity, likely generated by accretion onto a SMBH (Eisenhardt et al. 2012, Bridge et al. 2013, Jones et al. 2014, Tsai et al. in prep.) implied by their high dust temperatures. By comparing the mid-IR to submm luminosity ratios of Hot DOGs and other $z \sim$ 2 populations, Wu et al. (2012) suggested that SMGs, bump DOGs (DOGs with significant starburst contributions, based on mid-IR photometry, Dey et al. 2008), power-law DOGs (DOGs where the AGN contribution dominates, Dey et al. 2008) and Hot DOGs may follow an evolutionary sequence, driven by the growth of the central SMBH. Hot DOGs may be at a transitional phase between the DOG phase and the visible quasar phase, when gas and dust are being heated and blown out by powerful AGN feedback. Bridge et al. (2013) also propose this scenario, based on the discovery of a large fraction of extended Lyman $\alpha$ emission for WISE-selected galaxies identified by a related method.  

Near-IR to mm SEDs have been constructed to explore the nature of this rare population (e.g., Wu et al. 2012, Bridge et al. 2013, Jones et al. 2014, Tsai et al. in prep., Assef et al. in prep.). However, neither the WISE W3/W4 discovery images nor single-dish submm follow-up data have the necessary spatial resolution to permit unequivocal matches to optical/near-IR counterparts; higher-resolution submm/mm follow-up is necessary to pinpoint the SEDs of the longer wavelength sources. Submm/mm interferometric observations can also help to resolve high-redshift ULIRGs into multiple, associated systems, and reveal possible internal structure crucial to understanding the nature of the systems. For example, SMA (Iono et al. 2006, Wang et al. 2011b, Chen et al. 2011, Barger et al. 2012), ALMA (Karim et al. 2013), and IRAM (Walter et al. 2014)  follow-up of high-redshift SMGs have revealed several spuriously identified SMGs, and resolved some SMGs into small systems with multiple mergers. 

Here we report a pilot high-resolution study of three Hot DOGs. We obtained $\sim$ 1.3 mm continuum maps for WISE J223810.20+265319.7 (hereafter W2238+2653) using the Combined Array for Research in Millimeter-wave Astronomy (CARMA), and for WISE J014946.17+235014.5 (hereafter W0149+2350) using the 
Submillimeter Array (SMA).  We observed CO 3-2 and 4-3 transitions for WISE J181417.29+341224.8 (hereafter W1814+3412) and W0149+2350, respectively, using CARMA.  We also obtained high-resolution near-IR images for W2238+2653 with the W.M. Keck telescope, and for W0149+2350 with the {\it Hubble Space Telescope}, enabling comparison of their stellar component with the dust and gas distribution. 
Magnitudes in this paper are in the Vega system. Throughout this paper we assume a $\Lambda$CDM cosmology with  $H_0 = 71$ km s$^{-1}$ Mpc$^{-1}$, $\Omega_m =0.27$, and $\Omega_{\Lambda} = 0.73$. 

\section{Observations and Data Reduction} \label{obs}
\subsection{CARMA}
Table 1 lists the three Hot DOGs reported in this paper. They are among the first Hot DOGs identified by our program, which now includes over 100 spectroscopically confirmed sources (Eisenhardt et al. in prep.). Basic observing information is presented in Table 2. 

\begin{deluxetable}{lllcl}
\tabletypesize{\scriptsize}
\tablenum{1}
\tablewidth{4.5in}
\tablecaption{Target list}
\tablehead{
\colhead{Source} &
\colhead{R.A.} &
\colhead{Dec.} &
\colhead{$z$} &
\colhead{Ref.}}

\startdata
W0149$+$2350 & 01:49:46.17 & +23:50:14.5  &  3.228  &   Wu et al. (2012)  \\
W1814$+$3412 & 18:14:17.29 & +34:12:24.8  &  2.451   &  Eisenhardt et al. (2012)  \\    
W2238$+$2653 &  22:38:10.20 & +26:53:19.7   &   2.405    &  Eisenhardt et al. (in prep.) 
\\         
\enddata
\tablecomments{Coordinates are in the J2000 reference system. The uncertainty of the redshift for these three Hot DOGs is $\Delta z\sim 0.002$.  }
\end{deluxetable}

\begin{deluxetable}{lcccclc}
\tabletypesize{\scriptsize}
\tablenum{2}
\tablewidth{6.5in}
\tablecaption{Observing information}
\tablehead{
\colhead{Source} &
\colhead{Telescope} &
\colhead{Config} &
\colhead{Frequency} &
\colhead{Line/Cont} &
\colhead{UT Date} &
\colhead{Time on source}\\
\colhead{} &
\colhead{} &
\colhead{} &
\colhead{(GHz)} &
\colhead{} &
\colhead{} &
\colhead{(hr)}}

\startdata
W0149$+$2350 & SMA  &   extended &   218.52   &  cont.    &      2010 Sep 19, 20   &   8.5   \\
W0149$+$2350 & SMA  &   compact  &   218.52   &  cont.    &       2010 Nov 5     &   7.6   \\
W0149$+$2350 & CARMA &   D  &   109.04  &  CO 4-3   &        2011 Jun 20-22     &   8.9   \\
W1814$+$3412 &  CARMA &   D  &  100.17   &  CO 3-2      &      2011 Apr 13, May 25,  Jun 1  &  12.3   \\    
W2238$+$2653 &  CARMA   &  D  &  248.00    &     cont.    &     2011 Oct 6, 13, Nov 2 & 9.3  \\
\enddata
\end{deluxetable}

We observed W2238+2653 in $\sim$ 1.2 mm continuum on UT 2011 October 6, 13 and UT 2011 November 2, with the average opacities at 230 GHz of 0.32, 0.47 and 0.07, respectively. The receivers were tuned at a local oscillator frequency of 248 GHz, and the continuum was observed using contiguous $\sim$ 500 MHz bands (with $\sim$5 MHz channels) on the upper and lower correlator sidebands of all 15 antennas, for a total coverage of $\sim$ 3 GHz with our setting. We reduced CARMA continuum data for W2238+2653 with the software package Multichannel Image Reconstruction Image Analysis and Display (MIRIAD, Sault et al. 1995). We flagged and calibrated the visibility data using standard calibration procedures with MIRIAD. For each band we flagged the first two and last two frequency channels due to poor sensitivity at the bandpass edges. For data taken on UT 2011 October 6 and 13, we used MWC 349 as the flux calibrator, 3C84 as the passband calibrator and 2203+317 as the gain calibrator. For data taken on UT 2011 November 2, we used Uranus as the flux calibrator and 3C454.3 as the gain calibrator. The calibrated visibilities from the data sets were then combined. The imaging and deconvolution were done using MIRIAD tasks \textit{invert} and \textit{mossdi}. We used natural weighting to maximize the sensitivity to broad, faint structures. The \textit{mossdi} deconvolution task uses a Steer CLEAN algorithm (Steer et al. 1984).

The CARMA CO line data for W1814+3412 were taken on UT 2011 April 13, May 25 and June 1, with the average opacities at 230 GHz of 0.11-0,13. Only 7 antennas were used on April 13, while all 15 antennas were used on May 25 and June 1. We used Uranus as the flux calibrator, 3C454.3 as the passband calibrator, and 1801+440 as the gain calibrator for W1814+3412 observations. The CO data for W0149+2350 were taken on UT 2011 June 21-23, with the average opacities at 230 GHz of 0.15-0.25, and 15 antennas were used.  We used Uranus as the flux calibrator, 3C454.3 as the passband calibrator, and 0152+221 as the gain calibrator for W0149+2352 observations. All these CO observations used contiguous $\sim$ 500 MHz bands with $\sim$ 5 MHz channels. Individual source was phase, amplitude and passband calibrated using the standard routines in MIRIAD for each track. The files were then combined and the UV data were Fourier transformed and cleaned using the tasks \textit{invert} and \textit{mossdi} in MIRIAD. This produced a data cube with a resolution of 100 km s$^{-1}$.    

\subsection{The SMA}

SMA imaging of W0149$+$2350 was obtained as part of program
2010A-S052 (PI.  R.~S.~Bussmann). Observations were conducted in the compact array
configuration (maximum baseline length of $\sim 76~$m) on UT 2010 November 5 ($t_{\rm
int} = 7.6$~hr on-source integration time, $\tau_{\rm 225 \; GHz} \approx
0.08$).  Extended array observations (maximum baseline length of 226$\;$m) were
obtained on UT 2010 September 19 ($t_{\rm int} = 4.0$~hr on-source integration
time, $\tau_{\rm 225 \; GHz} \approx 0.09$) and on UT 2010 September 20 ($t_{\rm int}
= 4.5$~hr on-source integration time, $\tau_{\rm 225 \; GHz} \approx 0.10$).
The phase stability was good for all observing nights (phase errors between 10
and 30 degrees rms).

We used the SMA single-polarization 230~GHz receivers, tuned such that the CO ($J=8-7$) line
was covered at the center of the lower sideband. For W0149$+$2350 ($z = 3.228$),  we set the center frequency at 218.519~GHz for all observations. 
The SMA receivers provide an
intermediate frequency coverage of 4--8 GHz, totaling 8~GHz bandwidth
considering both sidebands.

Calibration of the {\it uv} visibilities was performed using the Interactive
Data Language (IDL) {\sc MIR} package.  The blazar 3C84 was used as the primary
bandpass calibrator and Callisto was used for absolute flux calibration.  The
nearby quasar 0237$+$288 ($F_{\rm 10\mu m} = 1.0~$Jy, 11 degrees from target)
was used for the phase gain calibration.  

The MIRIAD software package was used to invert the {\it uv}
visibilities and clean the dirty map.  Natural weighting provided maximum
sensitivity and resulted in an elliptical Gaussian beam with a full-width at half-maximum (FWHM) of $2\farcs24\times2\farcs08$ at a position angle of 80
degrees east of north for the combined compact and extended array image.

\subsection{Keck NIRC2}
W2238+2653 was observed with  NIRC2 behind the Keck II laser guide star (LGS) adaptive optics (AO) system (Wizinowich et al. 2006, Van Dam et al. 2006) on the night of UT 2011 August 16.  We used the $R$=14.2 USNO-B star 1168-0602706 (Monet et al. 2003) located 50$\arcsec$ from the target for the tip-tilt reference star.  Images were obtained with the MKO $Kp$ filter,  which is shifted to shorter wavelengths than the K or Kshort filters and has a slightly wider bandpass (Simons \& Tokunaga 2002); we used this filter because it is more sensitive due to lower thermal background. We used the wide camera setting (nominal pixel scale = 0\farcs0397/pixel) which provided a single-frame field of view of 40\arcsec$\times$40\arcsec.  Nine $Kp$-band images (120~sec per image) were obtained using a 3-position dither pattern that avoided the noisy, lower-left quadrant.  

The images were reduced using a custom set of IDL scripts.  The raw images were first dark-subtracted and then sky-subtracted using a sky frame created from the median average of all frames whose frame-level median value was less than a factor of 1.2 from that of the image being reduced.  Finally, a dome flat was used to correct for pixel-to-pixel sensitivity variations.  The reduced images were superposed on a larger array and shifted such that a star in common on all images was aligned to the same pixel location.  The aligned images were median averaged to form the final mosaic.

\subsection{\it Hubble Space Telescope} 

W0149+2350 was imaged for 2600 sec in the F160W filter with WFC3 on {\it HST} as part of program 12481 (PI C. Bridge) on UT 2011 November 7. The images were obtained at four dithered positions with the ``WFC3-IR-DITHER-BOX-MIN" pattern, using the MULTIACCUM mode with 14 samples 50 seconds apart at each position. The standard CALNICA and CALNICB pipeline from the Space Telescope Science Institute was used for image reduction.

\subsection{\it Far-IR and Submm data} 
 To constrain the cold dust temperatures of Hot DOGs detected in our CARMA and SMA observations, we used CSO 450 $\mu$m and {\it Herschel} 500 $\mu$m continuum measurements. The 450 $mu$m continuum data for W0149+2350 were taken using SHARC2 at the CSO, with details reported in Wu et al. (2012).   
The 500 $\mu$m continuum was observed by SPIRE on {\it Herschel} (PI: P. Eisenhardt, Proposal ID: OT1\_peisenha\_1) on UT 2011 January 3 and UT 2012 January 27 for W2238+2653 and W0149+2350, respectively. The SPIRE maps were made using small jiggle map mode, with a 487-second integration time per source. The data were processed and analyzed with HIPE v11.1.0. The complete {\it Herschel} observations for these and other Hot DOGs, with a more detailed analysis of the warm dust components, will be presented in Tsai et al. (in prep).

\section{Results} \label{results}

\subsection{Continuum}
In Table 3, we present the continuum measurements obtained from the CARMA and SMA continuum images for two Hot DOGs. 
W0149+2350 was well detected at 1.37 mm in both the compact and extended configuration of the SMA, with a flux density of 1.8$\pm$0.2 mJy in the combined image (Figure 1a). The source is unresolved in both the combined image and in the extended configuration image, with the highest resolution from the extended configuration of 1\farcs0. The flux densities derived from the compact and extended configurations are consistent within the calibration uncertainty, which is about 20\% for our SMA observations in addition to the uncertainties quoted in Table 3. This implies that the cold dust extends over less than 1\farcs0, or 7.6 kpc at $z$=3.228. The {\it HST} WFC3 image (F160W, effective wavelength 1.537 $\mu$m) of W0149+2350 is presented in Figure 1b and 1c. At the redshift of the source, F160W corresponds to rest-frame 3600~\AA, tracing the young stellar light of the host galaxy. The resolution of the {\it HST} WFC3 image is 0\farcs13, and the source is well resolved with an effective radius (Re, the radius within which half the light is enclosed) of 0\farcs34 (2.6 kpc). 

\begin{figure}
\epsscale{1.0}
\rotatebox{0}{\plotone{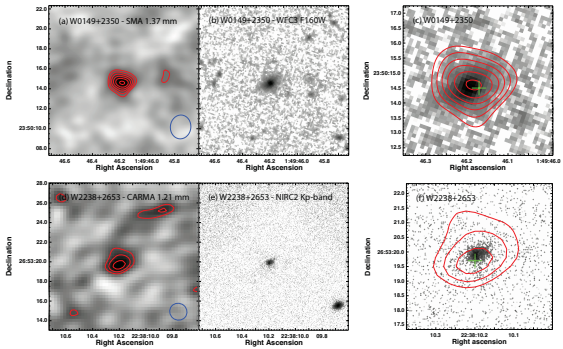}}
\caption{a) SMA image of W0149+2350 at 1.37 mm; the intensity contours start at the 3$\sigma$ level, with steps of 1 $\sigma$=0.22 mJy beam$^{-1}$. The beam size is shown at the lower right. b) {\it HST} F160W image of W0149+2350. c) A zoom-in {\it HST} F160W image of W0149+2350 overlaid with SMA 1.37 mm contours. The green cross marks the WISE position where the spectroscopic redshift was measured. d) CARMA image of W2238+2653 at 1.21 mm;  the intensity contours start at 3$\sigma$ level, with steps of 1 $\sigma$=0.84 mJy beam$^{-1}$. The beam size is shown at the lower right. e) Keck NIRC2 image of W2238+2653 in the $Kp$ band. f) A zoom-in NIRC2 image of W2238+2653 overlaid with CARMA 1.21 mm contours. The green cross marks the WISE position where the spectroscopic redshift was measured. Images cover $15\arcsec \times 15\arcsec$ in (a), (b), (d), (e) and $5\arcsec \times 5\arcsec$ in (c) and (f). } 
\end{figure}

\begin{deluxetable}{llcccccc}
\tabletypesize{\scriptsize}
\tablenum{3}
\tablewidth{6.5in}
\tablecaption{1.3 mm continuum measurements for Hot DOGs}
\tablehead{
\colhead{Source} &
\colhead{Telescope/Config} &
\colhead{Wavelength} &
\colhead{Beam size} &
\colhead{PA} &
\colhead{Flux density$^{a}$} &
\colhead{Peak position}\\
\colhead{} &
\colhead{} &
\colhead{(mm)} &
\colhead{} &
\colhead{(deg)} &
\colhead{(mJy)} &
\colhead{(J2000.0)}}

\startdata
W0149$+$2350 & SMA/Compact & 1.37 & 3\farcs4 $\times$ 3\farcs0 &80 & 1.8$\pm$0.2 & 01:49:46.18 +23:50:14.8 \\
W0149$+$2350 & SMA/Extended & 1.37 &  1\farcs3 $\times$ 1\farcs0 &80 & 2.3$\pm$0.4 & 01:49:46.19 +23:50:14.7 \\
W0149$+$2350 & SMA/Combine & 1.37 & 2\farcs2 $\times$ 2\farcs1 &80 & 1.8$\pm$0.2 & 01:49:46.19 +23:50:14.7 \\
W2338$+$2653 & CARMA/D & 1.21 & 1\farcs8 $\times$ 1\farcs7 &140 & 4.5 $\pm$0.8 & 22:38:10.20 +26:53:20.1 \\
\enddata
\tablenotetext{a}{The quoted uncertainties do not include in uncertainty in the calibration, which is about 20\% for SMA observations, and 15\% for CARMA observations.}
\end{deluxetable}

W2238+2653 was detected by CARMA at 1.21 mm with a flux density of 4.5$\pm$0.8 mJy (Figure 1d), with an additional 15\% uncertainty from calibration. The beam size of the image is 1\farcs8 $\times$ 1\farcs7 while the source size is $2\farcs9 (\pm 0\farcs3) \times 2\farcs6 (\pm 0\farcs3)$. Although it seems that the source size is about 50\% bigger than the beam and it is promising that the source is resolved, given the low S/N ratio ($\sim$ 5), we can not be very certain about its extent based on the CARMA data alone. W2238+2653 was also detected by Bolocam at CSO at 1.1 mm with a flux density of 6.0$\pm$2.2 mJy (Wu et al. 2012). Assuming $\beta$=1.5 and T$_d$=35K (see Section 4.2 for why these parameters were chosen), this corresponds to a flux density of 4.6$\pm$1.7 mJy at 1.21 mm within the 42$\arcsec$ Bolocam beam, essentially identical to the  4.5$\pm$0.8 mJy CARMA value shown in Table 3. The consistency between the CARMA and Bolocam flux densities (within their uncertainties) suggests that not much emission comes from the region outside the $\sim 2\arcsec$ CARMA beam. 
Figure 1e and 1f present the NIRC2 image of W2238+2653 in the $Kp$ band, centered at 2.124 $\mu$m. This wavelength corresponds to rest frame 6230~\AA, tracing old stellar light from the host galaxy. The NIRC2 image resolution is 0\farcs16, and the effective radius of the source is 0\farcs42 (3.4 kpc). 

Our CO observations of W0149+2350 and W1814+3412 with CARMA also obtained continuum at $\sim$ 3 mm. However, since the flux density of these Hot DOGs drops dramatically from 1 mm to 3 mm, and the CO observations were designed to optimize line detection and thus have limited bandwidth, the noise obtained from these 3 mm continuum observations is much higher than predicted flux density.  As expected, neither source was detected in the 3mm continuum, with 2$\sigma$ upper limits of 2.2 mJy and 1.2 mJy for W0149+2350 and W1814+3412, respectively. The predicted 3 mm continuum flux densities (assuming $\beta$=1.5 and T$_d$=35K, see Section 4.2) based on the 1.3 mm SMA detection for W0149+2350 and the Bolocam 1.1 mm upper limits for W1814+3412 (Wu et al. 2012) are 0.11 mJy and less than 0.07 mJy for W0149+2350 and W1814+3412, respectively, which are an order of magnitude lower than the observed limits. 

For both W2238+2653 and W0149+2350, the peak position in the SMA and CARMA images agrees well with the WISE position (see Figure 1c and 1 f, differences are less than 0\farcs4) where the redshift is measured, confirming that the high luminosity source coincides with the WISE-selected target. From the {\it HST} and Keck AO images and the submm interferometric maps, there is no clear morphological evidence for strong gravitational lensing, in which case the source would likely have a multiple, offset, or arc-like distortion set of light peaks, rather than the simple centrally concentrated and roughly axisymmetric features we see at all wavelengths (Figure 1).  These are in marked contrast with known and confirmed strongly lensed star-forming galaxies. Examples include
candidates selected from SDSS data and confirmed by HST observations (e.g.
Bolton et al. 2008), and candidates selected by  South Pole Telescope observations and confirmed by ALMA (Hezaveh et al. 2013). At the sensitivity of the data we report on here, a configuration suggestive of strong lensing behavior would have been evident.

\subsection{Molecular gas}

We attempted to detect CO 3-2 for W1814+3412, and CO 4-3 for W0149+2350 at 3 mm using CARMA in the D configuration. Both targets were undetected, with the implied 1$\sigma$ CO (4-3) limits of 0.85 Jy km s$^{-1}$ for W0149+2350 and CO (3-2) limits of 0.46 Jy km s$^{-1}$ for W1814+3412. The CO (8-7) transition was also covered by the 1 mm continuum observation of W0149+2350 using the SMA. 
No detection was obtained on this high-J transition, with a 1$\sigma$ CO (8-7) limit of 0.45 Jy km s$^{-1}$. Although two J ladder transitions (CO 4-3 and CO 8-7) were observed for W0149+2350, its CO excitation can not be further constrained since both lines were undetected. In order to estimate the molecular gas traced by the CO (1-0) luminosity, we used empirical conversion factors from  Carilli \& Walter (2013). These are 0.97 for CO (3-2) to CO (1-0), and 0.87 for CO (4-3) to CO (1-0). These luminosity ratios are for quasar-like targets, which we adopted since Hot DOGs are dominated by strong AGNs. The conversion factors are lower for SMGs: 0.66 for CO (3-2) to CO (1-0), and 0.46 for CO (4-3) to CO (1-0), leading to up to a factor of two lower CO luminosity and gas mass estimate, and can be considered a lower limit. Higher J CO transitions like CO (8-7) generally do not contribute significantly to the total CO luminosity for High-z quasars and SMGs (e.g. Wei{\ss} et al. 2007b), except in unusual cases like APM J08279+5255 (e.g. Wei{\ss} et al. 2007a), where the CO ladder SED peaks at J=10. Although we know little about the excitation for these Hot DOGs, the non-detection of W0149+2350 in CO (8-7) at a lower limit than the non-detection in CO (4-3) suggests it is unlikely to be a very high excitation AGN like APM J08279+5255. Therefore, in this paper we only use the CO (4-3) limit to estimate the gas mass limit for W0149+2350.

To calculate the upper limit of molecular gas mass from the CO measurement, we assumed a FWHM linewidth of 400 km s$^{-1}$, the median value for detected CO lines in $z >$1 quasars (e.g. Carilli \& Walter 2013).  Taking the approach used in Solomon et al. (1997), the line luminosity of CO is given by \begin{equation}
L'_{\rm CO}=3.25\times 10^{7}\times I_{\rm CO} \nu_{\rm obs}^{-2} D_{L}^{2} (1+z)^{-3}, \end{equation}
where $L'_{\rm CO}$ is in units of ${\rm K~km~s^{-1}~pc^{-2}}$,  $I_{\rm CO}$ is the integrated intensity of the CO line in Jy km s$^{-1}$, $\nu_{\rm obs}$ is the observing frequency in GHz, and $D_{L}$ is the luminosity distance in Mpc. To convert CO (1-0) luminosity to total molecular gas mass $M_{H_{2}}$, an empirical relation is taken: 
$M_{H_{2}}=\alpha L'_{\rm CO}$, where $M_{H_{2}}$ is in units of $M_{\odot}$ and $L'_{\rm CO}$ is in units of ${\rm K~km~s^{-1}~pc^{-2}}$.  The converting factor $\alpha$ has a typical value of 0.8 for bursting galaxies, and a much higher value ($\sim$ 4) for main sequence galaxies (e.g. Bolatto, Wolfire, and Leroy 2013, Carilli \& Walter 2013). We use $\alpha = 0.8$ for Hot DOGs in this paper. The derived parameters are listed in Table 4. Another attempt to measure CO and resolved continuum emission at observed frame 1 mm for W1814+3412 was conducted using the IRAM telescope, which is reported in Blain et al. (in prep.).

\begin{deluxetable}{llccccc}
\tabletypesize{\scriptsize}
\tablenum{4}
\tablewidth{6.5in}
\tablecaption{CO measurements for Hot DOGs}
\tablehead{
\colhead{Source} &
\colhead{Line} &
\colhead{1$\sigma$ rms} &
\colhead{rms bin} &
\colhead{$I_{\rm CO}$ limit (1$\sigma$)} &
\colhead{$L'_{\rm CO}$  limit (2$\sigma$)} &
\colhead{$M_{\rm H_{2}}$ limit (2$\sigma$)}\\
\colhead{} &
\colhead{} &
\colhead{(mJy/beam km s$^{-1}$)} &
\colhead{(km s$^{-1}$)} &
\colhead{(Jy km s$^{-1}$)} &
\colhead{(10$^{10}$K km s$^{-1}$ pc$^{-2}$)} &
\colhead{(10$^{10}$M$_{\odot}$)}}

\startdata
W0149$+$2350 & CO 4-3 & 4.3 & 100 &  0.85 & $<$  4.8  & $<$  3.3  \\  
W1814$+$3412 & CO 3-2 & 2.3 & 100 &  0.46 & $<$  3.0 & $<$  2.3 \\    
\enddata
\end{deluxetable}

\section{Discussion } \label{sourcedisc}

\subsection{Morphology}
For both W2238+2653 and W0149+2350 only a single, centrally condensed source is seen in the near-IR and mm images, with no obvious signatures of merging such as multiple cores, tails or arms. The consistent flux densities from both the compact and extended configurations for W0149+2350, and from the CARMA and CSO observations for W2238+2653, suggest neither a halo of faint emission, nor do fainter compact sources seem to be contributing substantially. This argues that these two sources are single and compact  (Here we refer sources as compact when their  angular size is less than 2$\arcsec$). If these Hot DOGs do have a merging origin,  and if the theory of galaxy evolution through merging (e.g. Hopkins et al. 2008) and a galaxy evolutionary sequence of SMGs-DOGs-Hot DOGs-quasars suggested in Wu et al. (2012) is true, then the morphology of these two Hot DOGs support the scenario that the merging event was not in the recent past ( i.e. $\sim 80\,$Myr, Mihos \& Hernquist 1996), and the Hot DOGs are likely closer to the quasar phase. We use the Sersic (1968) index to  explore if the sources have a disk-type or bulge-type light profile (e.g., Simmons \& Urry 2008). Using GALFIT (Peng et al. 2002), we obtained Sersic indices $n$ from the NIRC2 image and {\it HST} images. The Sersic index is 1.4 for W2238+2653, and 1.6 for W0149+2350, indicating a disk-type morphology. These values are similar to the typical $n < 2$ found for DOGs (Bussmann et al. 2009, 2011). 

About one third of optically selected quasars have submm/mm continuum detections at mJy levels out to high redshift (Priddey et al. 2003, Beelen et al. 2006, Wang et al. 2008), and most of the submm-detected quasars yield CO detections. The size of their CO reservoirs are typically compact, less than a few kpc (e.g., Walter et al. 2004, Riechers et al. 2011). These gas-rich quasars are thought to correspond to an earlier phase of activity than submm-faint quasars. SMG galaxies, however, have been found to have larger cold gas reservoirs, typically larger than $\sim$ 10 kpc  (e.g., Ivison et al. 2010, Carilli et al. 2011, Hodge et al. 2012). Based on our 1.3 mm images, the cold dust size is $<$ 8 kpc in diameter for W0149+2350, consistent with the size of the stellar emission in near-IR image. The cold dust may be more extended for W2238+2653,  but the size of dust emission is quite uncertain due to the low S/N CARMA detection, and the agreement between the 2$\arcsec$ CARMA flux density and that from the 42$\arcsec$ CSO Bolocam data argues against significant emission beyond a 16 kpc diameter. The half light radius in the near-IR image of  W2238+2653 is 3.4 kpc. (Without a much higher S/N submm interferometric observation, we can not meaningfully compare the cold dust extent to the stellar extent). These two cases suggest the size of Hot DOGs is compact compared to SMGs and likely similar to gas-rich quasars.

\subsection{Mass of cold dust for Hot DOGs}
The most striking characteristic feature of Hot DOGs is that their SEDs have unusually strong mid-IR emission compared to their submm emission, implying hot dust dominates the energy output of these objects. Wu et al. (2012) found that despite the much stronger mid-IR luminosity (presumably due to AGN heating) for Hot DOGs than for SMGs and standard DOGs, the cold dust emission traced by 350$\mu$m is actually not very different for these populations. If we simplistically assume that there are two major dust components in these galaxies---an AGN-heated hot dust component and a starburst-heated colder dust component---then the similar submm emission levels of Hot DOGs, SMGs and DOGs suggest that these populations share a similar level of star formation activity.

In reality, these systems are likely more complicated than this simple two-temperature dust component model, but our longest wavelength data tends to trace the coldest major dust component, which is less likely to be heated by an AGN, and more likely to be heated by star formation. Wu et al. (2012) used data at 350 $\mu$m and 1.1 mm to constrain the cold dust properties of Hot DOGs, finding that they are not very different from local starburst galaxies like Arp220 (a local analog to SMGs) and the AGN/starburst composite system Mrk231 (a good SED template for standard power-law DOGs; e.g., Bussmann et al. 2009b,  Melbourne et al. 2012). In addition to the hotter component, with dust temperatures ranging up to several hundred K, Hot DOGs likely also have a starburst-related cold dust component with a temperature between 30-50 K, as suggested by their 350 $\mu$m to 1.1 mm ratios (Wu et al. 2012). The real typical temperature could be closer to 30 K than 50 K considering that 350 $\mu$m data could be affected more by the AGN heating than the 1.1 mm data, and such a temperature is similar to cold dust temperatures found in SMGs and DOGs (e.g. Magnelli et al. 2012, Melbourne et al. 2012). The presence of a mm-wave continuum limit allows the possible combinations of temperature and mass in such a cold dust component to be constrained.   

Cold dust mass is a key parameter for understanding galaxies, providing insight into gas and thus total baryonic mass via assumed conversions. The SEDs of Hot DOGs are flat (in units of $\nu f_{\nu}$) from mid-IR to submm wavelengths (Wu et al. 2012, Bridge et al. 2013, Jones et al. 2014, Tsai et al. in prep.), implying the AGN-heated mid-IR component is of comparable strength to the starburst-related submm component in terms of energy output. 
Since energy output goes as the 4th power of temperature at short wavelengths where the dust emission is optically thick, and which contributes significantly to the near-to-mid IR SED, the colder dust ($T_d < 50$ K) associated with star formation constitutes a significant fraction of the overall dust mass.
We can roughly estimate the mass of the cold dust using the longest wavelength data and a single temperature blackbody model:
\begin{equation}
M_d= \  \frac{D_{L}^{2}}{(1+z)} \cdot \frac{S(\nu_{\rm obs})}{
\kappa(\nu_{\rm rest})B(\nu_{\rm rest},T_d)}\end{equation} 
where $M_d$ is the dust mass, $D_L$ is the luminosity distance, $\nu_{\rm rest}$ is the rest frequency of the observed band (which are 925 GHz and 824 GHz for W0149+2350 and W2238+2653, respectively), $B_{\nu}$ is the Planck function, $\kappa_{\nu} =
\kappa_{0}(\nu/1 {\rm THz})^{\beta}$ is the dust mass absorption coefficient, and $\beta$ is the emissivity index. In this paper, we adopt $\kappa_{0}$ = 2.0\  m$^{2}$ kg$^{-1}$ (Draine et al. 2003), and assume a dust temperature of 35 K (roughly the median dust temperature found in SMGs and DOGs; Magnelli et al. 2012, Melbourne et al. 2012) and $\beta$=1.5 (typical emissivity index assumptions for SMGs and DOGs).

Based on the $\sim$1.3 mm continuum measurements, we calculate cold dust masses of 2.0$\times 10^{8} M_{\odot}$ for W0149+2350 and 3.9$\times 10^{8} M_{\odot}$ for W2238+2653. However, there are large uncertainties in these estimates. Besides the $10-20$\% statistical flux density uncertainties, we also need to consider uncertainties in $\kappa$ and T$_d$. Changing $\beta$ from 1.5 to 2.0 only changes $\kappa$ by less than 10\%; however, $\kappa_{0}$ is uncertain to at least a factor of two (Dunne et al. 2003).  Dust temperature uncertainties can bring in another a factor of two or more uncertainty, and likely dominate the overall uncertainty in dust mass estimates. 

To constrain the cold dust temperature, we set upper limits on the cold dust component temperature  (and lower limits on the cold dust mass) for W0149+2350 and W2238+2653, by using the longest wavelength (500$\mu$m) continuum data available from {\it Herschel}. W2238+2653 was detected at 500 $\mu$m by SPIRE on {\it Herschel}, with a flux density of 61.2$\pm$6.0 mJy. While no {\it Herschel} detection was made of W0149+2350 ($3 \sigma$ upper limit of 57 mJy), we detected a flux density of 35$\pm$9 mJy at 450 $\mu$m using SHARC2 at the CSO (Wu et al. 2012). Using the flux density ratios of the 450$\mu$m or the 500 $\mu$m measurements to the $\sim$1.3 mm measurements, we estimate upper limits for the cold dust temperatures with a single-temperature blackbody model, adopting $\beta$=1.5. The derived temperature  limits are noticeably warmer: 55 K for W0149+2350, and 69 K for W2238+2653, corresponding to  cold dust masses of 1.1$\times 10^{8} M_{\odot}$ and 1.4$\times 10^{8} M_{\odot}$ for W0149+2350 and W2238+2653, respectively. 

These derived temperatures by including both 450/500 $mu$m and 1.3 mm data are higher than the representative temperatures of SMGs and DOGs. Since the mid-IR to submm ratio is unusually high in these Hot DOGs, it is plausible that even continuum at 450-500 $\mu$m (rest-frame 100-150 $\mu$m) is significantly affected by the AGN contribution, and {\it Herschel} data does not trace the star formation-related cold dust well as it does in typical SMGs. Checking continuum data for Hot DOGs in Wu et al. (2012) that have both 350$ \mu$m and 1.1 mm detections, we find a similar trend, that including both 350 $\mu$m and 1.1 mm measurements, the derived dust temperature are higher, and the derived cold dust masses are generally a factor of two  smaller than those derived from 1.1 mm emission alone using the same single-temperature blackbody model. This implies a more significant contribution from AGN heating to the 350 $\mu$m emission than to the 1.1 mm continuum. For this reason, we consider the above cold dust mass estimates including the $\sim$450-500$ \mu$m data as a lower limit. To constrain the upper limit of the cold dust mass, we take a lower limit of the dust temperature as 20 K, which is about the average dust temperature of the Milky Way (Wright et al. 1991). Assuming $\beta$=1.5, this corresponds to cold dust mass upper limits of 7.0$\times 10^{8} M_{\odot}$ and 1.2$\times 10^{9} M_{\odot}$ for W0149+2350 and W2238+2653, respectively. A more accurate way to estimate the star-formation related cold dust mass is to obtain multiple measurements at a range of wavelengths longer than 1 mm and fit the cold dust component. Alternatively, one can construct a complete mid-IR to mm SED and decompose the different dust components with different temperatures, separating contributions from the AGN and starburst, as a recent example shown in Drouart et al. (2014). This is beyond the scope of this paper, and could be subject to free-free contamination at very long wavelengths.

Although simplistic in nature, our approach to estimate cold dust masses based on the continuum measurement at $\sim$ 1 mm allows to compare Hot DOGs with other z $\sim$ 2  bursting galaxy populations (SMGs, DOGs, quasars, radio galaxies), as shown in Figure 2. Longer wavelength detections were reported in Wu et al. (2012) for six additional Hot DOGs at 1.1mm, and in Jones et al. (2014) for six hot DOGs at 850 $\mu$m. We included longer wavelength detections for other bursting galaxy populations, mostly from Carrili \& Walter (2013) and references therein.  We use a consistent dust temperature T$_d$=35 K and emissivity $\beta$=1.5 for all calculations. 
Known lensed targets have been excluded from our sample. 

\begin{figure}
\epsscale{1.0}
\rotatebox{0}{\plotone{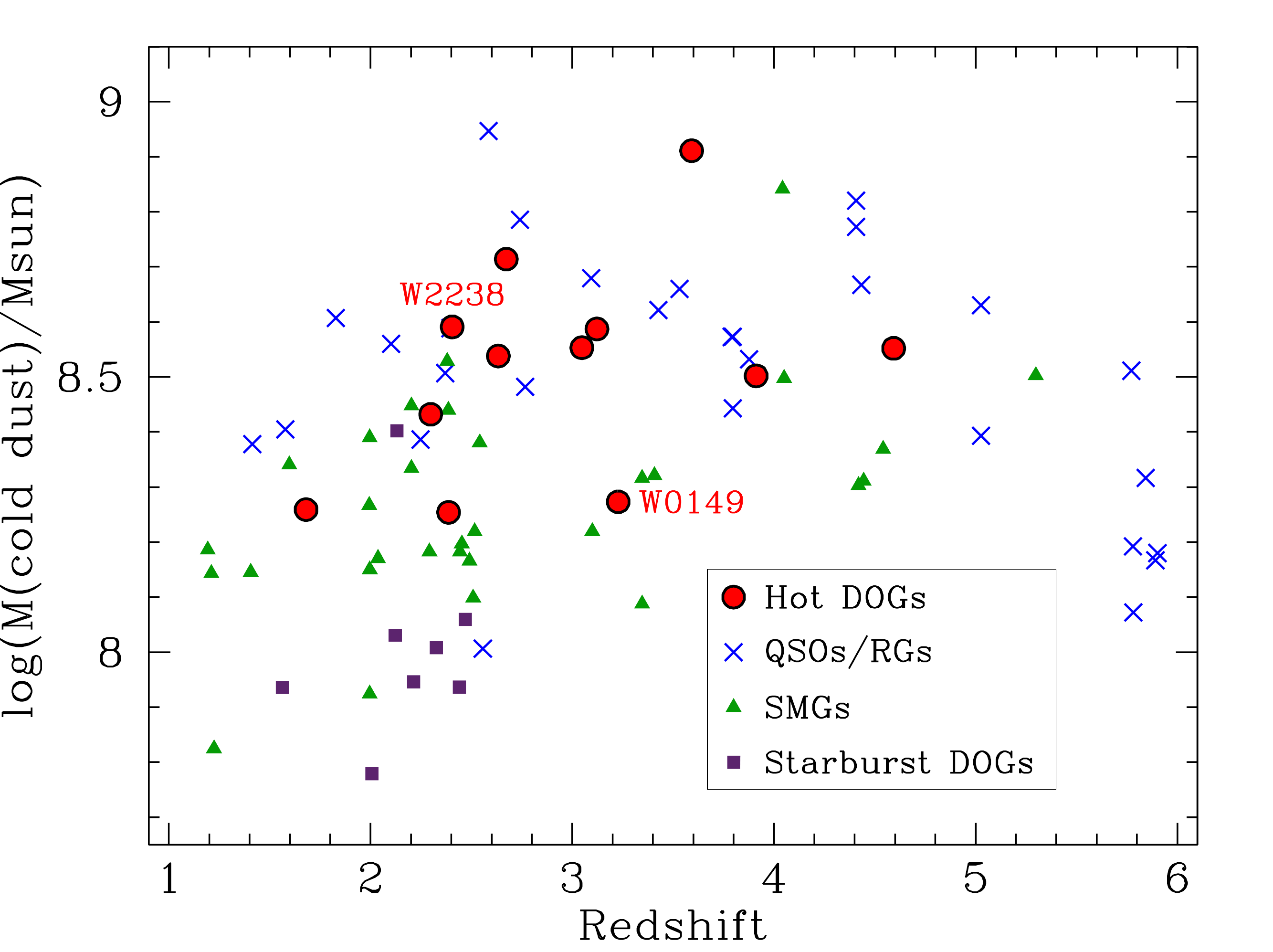}}
\caption{Cold dust masses calculated assuming $\beta$=1.5 and $T_d$=35K from $\sim$1 mm continuum detections for Hot DOGs (this work, Wu et al. 2012, Jones et al. 2014) and QSOs/RGs, starburst DOGs and SMGs (Carilli \& Walter 2013 and references therein). 
}
\end{figure}

The cold dust masses of Hot DOGs are comparable to those in submm-detected quasars and radio galaxies, implying they are both massive populations. This may be a hint of a possible evolutionary connection between Hot DOGs and visible quasars: Hot DOGs may be the progenitors of very luminous quasars, but are still deeply embedded in a hot dust cocoon, which obscures the AGN at optical/near-IR wavelengths (Assef et al. in prep.) and X-ray wavelengths (Stern et al. 2014). The SMGs in Figure 2 are about a factor of two less massive than quasars and Hot DOGs. However, considering the large uncertainty due to dust temperature, and that both quasars and Hot DOGs host powerful AGNs as an additional heating source, this difference in cold dust mass may not be significant. High-redshift main sequence galaxies are generally very faint at $\sim$ 1 mm (e.g., Magnelli et al. 2012b), implying that their derived cold dust masses are generally smaller, or their cold dust temperatures are lower than the bursting galaxies in Figure 2.

\subsection {Molecular gas for Hot DOGs}
Molecular gas probed by CO is another important measurement of total mass for starburst galaxies.  Although we did not obtain detections, we compare the $2\sigma$ upper limit on gas mass for the two Hot DOGs to detected molecular gas mass in other populations in Figure 3. The detected molecular gas mass of SMGs, DOGs, quasars and RGs plotted in Figure 3 are collected from public references (see caption for references). Their gas masses are directly quoted as reported in the references, which were all converted from solid CO detections, with proper CO excitation models applied to convert to CO 1-0 luminosity, and adopted  converting factor of 0.8 from CO luminosity to molecular gas mass. In Figure 3, the gas mass of SMGs and quasars are comparable, while starburst DOGs are slightly less massive. Our upper limits suggest that these two Hot DOGs might not be as gas-rich as the quasars in Figure 3. However, due to the limited number of Hot DOGs reported here and the shallowness of our CARMA data, we do not yet have strong constraints on the molecular gas content of Hot DOGs. Longer integrations and/or larger interferometer arrays (e.g., the Atacama Large Millimeter/sub-millimeter Array, ALMA) are needed to better constrain the CO gas mass of Hot DOGs. 

\begin{figure}
\epsscale{0.80}
\rotatebox{270}{\plotone{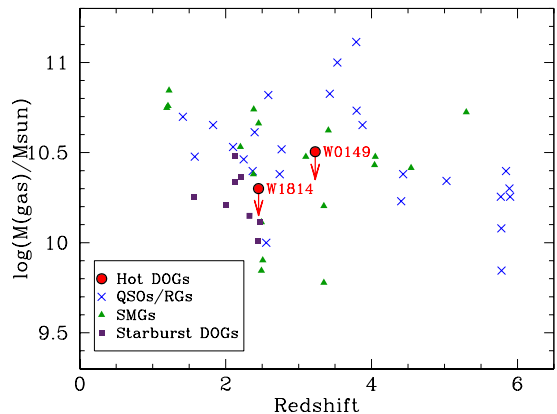}}
\caption{Upper limits (2 $\sigma$) for molecular gas mass for the two Hot DOGs based on CO measurements with CARMA, compared to molecular gas masses of other high-redshift populations with CO detections (QSO/RGs: Guilloteau et al. 1997, Carilli et al. 2002, 2007,  Greve et al. 2004, de Breuck et al. 2005,  Krips et al. 2005, Maiolino et al. 2007, Willott et al. 2007, Coppin et al. 2008, Wang et al. 2010, 2011, Polletta et al. 2011, Ivison et al. 2012, Schumacher et al. 2012; SMGs: Neri et al. 2003, Greve et al. 2005, Tacconi et al. 2006,  Frayer et al. 2008, Schinnerer et al. 2008, Daddi et al. 2009, 2009b, Engel et al. 2010, Riechers et al. 2010; Starburst DOGs: Yan et al. 2010). 
}
\end{figure}

\section{Summary} \label{summary}
We have obtained pilot high-resolution continuum and CO follow-up of three WISE-selected hot dust-obscured galaxies (Hot DOGs). The main results of our observations are as follows:

1. W0149+2350 is detected with a flux density of 1.8 $\pm$ 0.2 mJy at 1.37 mm using the SMA, and W2238+2653 with a flux density of 4.5 $\pm$ 0.8 mJy at 1.21 mm using CARMA. The data suggest both sources are single and compact.  W0149+2350 is unresolved, implying a source size less than 8 kpc in diameter. The consistency of the CARMA and CSO flux densities for W2238+2653 implies a source size less than 16 kpc in diameter.

2. W2238+2653 is well resolved in near-IR adaptive optics imaging with the W.M. Keck telescope.  W0149+2350 is well resolved in near-IR imaging with the {\it Hubble Space Telescope}. Their light profiles reveal single, centrally condensed sources with effective radii less than 4 kpc. 

3. There is good agreement between the near-IR and mm peak positions, confirming the high luminosity source coincides with the WISE-selected targets. 

4. There is no obvious morphological evidence for strong gravitational lensing.

5. The estimated mass of star formation-related cold dust is 2.0$\times 10^{8} M_{\odot}$,  for W0149+2350, with a range of 1.1$\times 10^{8} M_{\odot}$ to 7.0$\times 10^{8} M_{\odot}$. For W2238+2653 we estimated a cold dust mass of 3.9$\times 10^{8} M_{\odot}$,  with a range of 1.4$\times 10^{8} M_{\odot}$ to 1.2$\times 10^{9} M_{\odot}$. The cold dust mass of Hot DOGs is comparable to that of luminous quasars and SMGs.
 
6. We observed CO 3-2 and 4-3 transitions for W1814+3412 and W0149+2350, respectively, using CARMA. No detections were made. The corresponding 2$\sigma$ upper limits of molecular gas masses are 3.3$\times 10^{10} M_{\odot}$ and  2.3$\times 10^{10} M_{\odot}$ for W0149+2350 and W1814+3412, respectively.

\acknowledgements
This publication makes use of data products from the Wide-field Infrared Survey Explorer, which is a joint project of the University of California, Los Angeles, and the Jet Propulsion Laboratory/California Institute of Technology, funded by the National Aeronautics and Space Administration. This work is based on observations made with the Combined Array for Research in Millimeter-wave Astronomy. Support for CARMA construction was derived from the states of California, Illinois, and Maryland, the James S. McDonnell Foundation, the Gordon and Betty Moore Foundation, the Kenneth T. and Eileen L. Norris Foundation, the University of Chicago, the Associates of the California Institute of Technology, and the National Science Foundation. Ongoing CARMA development and operations are supported by the National Science Foundation under a cooperative agreement, and by the CARMA partner universities.  This work is also based on observations made with  the Submillimeter Array. The Submillimeter Array is a joint project between the Smithsonian Astrophysical Observatory and the Academia Sinica Institute of Astronomy and Astrophysics and is funded by the Smithsonian Institution and the Academia Sinica. Some of the data are based on observations made with the NASA/ESA {\it Hubble Space Telescope}, obtained at the Space Telescope Science Institute, which is operated by the Association of Universities for Research in Astronomy, Inc., under NASA contract NAS 5-26555. Some of the data presented herein were obtained at the W.M. Keck Observatory, which is operated as a scientific partnership among Caltech, the University of California and NASA. The Keck Observatory was made possible by the generous financial support of the W.M. Keck Foundation. This work uses data from {\it Herschel}. {\it Herschel} is an ESA space observatory with science instruments provided by European-led Principal Investigator consortia and with important participation from NASA. This work uses data from CSO, which is operated by the California Institute of Technology under funding from the National Science Foundation, contract AST 90-15755. R.J.A. was supported by Gemini-CONICYT grant number 32120009.


\end{document}